\documentclass[%
superscriptaddress,
amsmath,amssymb,
reprint,dyn
nofootinbib,
]{revtex4-2}
\usepackage{graphicx}
\usepackage{caption}
\usepackage{subcaption}
\usepackage{hyperref}
\usepackage{siunitx}
\usepackage{cleveref}
\usepackage{dcolumn}

\usepackage[mathlines]{lineno}
\modulolinenumbers[5]
\usepackage{hyperref}
\hypersetup{
  colorlinks=true,
  citecolor=blue,
  linkcolor=blue,
  urlcolor=blue
}

\begin{document}

\title{Measurement of Smartphone Sensor Efficiency to Cosmic Ray Muons}
\author{Jeff Swaney}
\affiliation{Department of Physics and Astronomy, University of California, Irvine, CA 92697, USA}
\author{Michael Mulhearn}
\affiliation{Department of Physics and Astronomy, University of California, Davis, CA, USA}
\author{Christian Pratt}
\affiliation{Department of  Physics and Astronomy, University of California, Davis, CA, USA}
\author{Chase Shimmin}
\affiliation{Department of Physics, Yale University, New Haven, CT, USA}
\author{Daniel Whiteson }
\affiliation{Department of Physics and Astronomy, University of California, Irvine, CA 92697, USA}

\begin{abstract}
A measurement of the efficiency of CMOS sensors in smartphone cameras to cosmic ray muons is presented.   A coincidence in external scintillators indicates the passage of a cosmic ray muon, allowing the measurement of the efficiency of the CMOS sensor. The observed flux is consistent with well-established values, and efficiencies are presented as a function of the number of photo-electrons collected from the CMOS silicon photodiode pixels. These efficiencies are vital to understanding the feasibility of large-scale smartphone networks operating as air-shower observatories.

\end{abstract}

\maketitle

\section{Introduction}
\label{sec:Intro}

The world-wide network of billions of consumer smartphones with CMOS-based cameras represent a largely untapped scientific resource.  The CMOS sensors are sensitive to ionizing radiation~\cite{2002SPIE.4669..172S}, and together with their processing and networking capabilities can operate as a network of particle detectors.  Such a network could be used to detect sources of radiation~\cite{Cogliati:2014uua,KANG2016126}, facilitate educational programs~\cite{decoaps}, or reconstruct extensive air showers from  astrophysical particles~\cite{bib:2.65} of unknown origin.

Ultra-high energy cosmic rays (UHECR), particles incident upon Earth's atmosphere with energy above \SI{e18}{eV}, are an enduring mystery. Despite many theoretical conjectures~\cite{Bell:1978zc,Blandford:1987pw,Waxman:1995vg,Weiler:1997sh}, their astrophysical source remains unexplained~\cite{Abraham:2007si,Abraham:2007bb}.  Such high-energy particles produce extensive air showers, which are detectable via their particle flux on the ground, fluorescence in the air, or radio and acoustic signatures.

Large, specialized facilities~\cite{Abbasi:2002ta,Takeda:1998ps,Abraham:2008ru,Aab:2014ila,AbuZayyad:2012ru,Hayashida:1998qb} observe cosmic
rays, reporting energies up to  $3\cdot
10^{20}$ eV. Such high-energy observations are rare and precious due to interaction with the cosmic microwave
background~\cite{Greisen:1966jv,Zatsepin:1966jv} which suppresses particle flux precipitously above $10^{18}$ GeV. Accumulating larger samples of UHECR using traditional mechanisms would require longer observations or larger facilities. 

Ref.~\cite{bib:2.65} explored the feasibility of a  strategy which employs the world-wide network of billions of consumer smartphones to detect the passage of air-showers through deposition of energy in the CMOS sensors which are the basis of smartphone cameras. A global network could complement the observation power of existing facilities, as well as be sensitive to globally-correlated pairs of showers from novel phenomena~\cite{Albin:2021psb}.  The sensitivity of such a network relies on the number and density of participating users, but also on the efficiency of smartphones to the air shower's muons and photons.  The authors of Ref.~\cite{bib:2.65} used radioactive sources to measure the efficiency to photons, and demonstrated muon sensitivity in a beam test, but lacked a precise measurement of the CMOS muon efficiency.

In this paper, the first dedicated measurement of the CMOS muon efficiency for several representative phone models is presented, using cosmic ray muons  tagged with coincidences in adjacent scintillator panels. Our results are expressed as a function of a threshold on the number of photo-electrons collected from the CMOS's silicon photodiode pixels, to allow comparison with models of energy deposition. This requires determining the internal gain, which is inferred from the Poisson nature of the pixel statistics.  Our measurement is directly applicable to estimates of the scientific power of smartphones to detect cosmic ray muons, as it is determined from the cosmic ray flux itself. 

\section{Experimental Design}

\subsection{Geometry}

\begin{figure}[h!]
    \centering
    \includegraphics[width=0.4\textwidth]{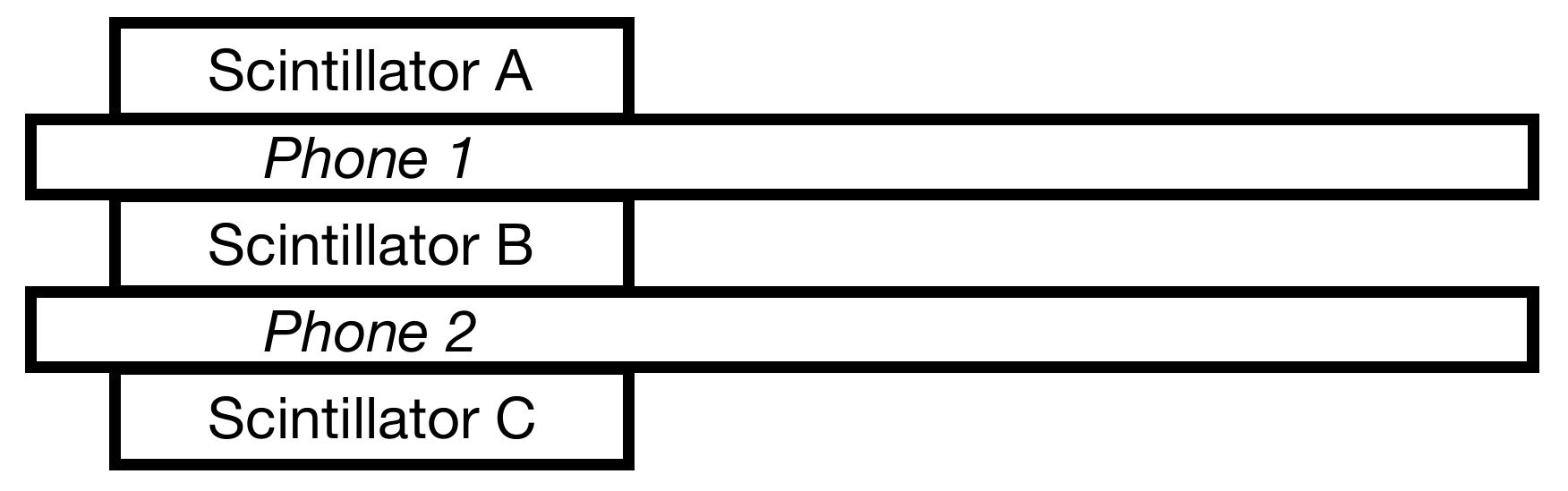}
    \caption{Cartoon schematic of the experiment.}
        \label{fig:schema}
\end{figure}

Two phones are stacked between three panels of scintillators, illustrated in \Cref{fig:schema}.

\noindent A coincidence in two scintillators serves as a {\it tag}, indicating the passage of a muon, which is used to measure the response of the smartphone CMOS sensor between them. The strategy of the experiment is  to assemble a set of cosmic muons tagged via coincidences in  adjacent scintillators; from this nearly-pure sample, the muon  efficiency of the CMOS sensors is measured as a function of a threshold on the per-pixel response. Though only a single phone between two scintillators is necessary, this configuration allows for more efficient running and for a valuable geometrical cross-check: the outermost scintillators (A and C) are also used to tag muons, though with a significantly lower acceptance.

\begin{figure}
    \centering
    \includegraphics[trim=50 0 80 0, clip, angle=270, origin=c, width=7cm]{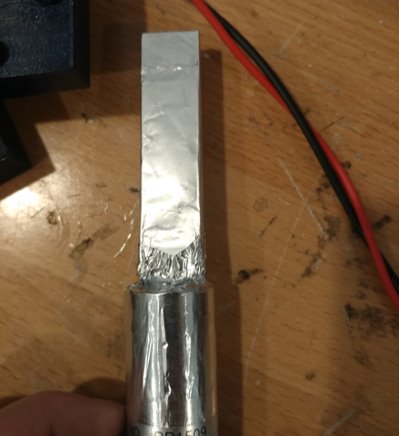}
    \vspace{-50pt}
    \caption{A photograph of the foil-wrapped photomultiplier tube, waveguide, and assembly.}
    \label{fig:pmt_al}
\end{figure}

The LYSO scintillators measure 16×14×8 mm, and are attached to photomultiplier tubes (PMTs) via a waveguide. To shield against light contamination, each scintillator, waveguide, and PMT are wrapped in multiple layers of aluminum foil and secured with foil tape; see \Cref{fig:pmt_al}. 

\begin{figure}
    \centering
    \includegraphics[height=2cm]{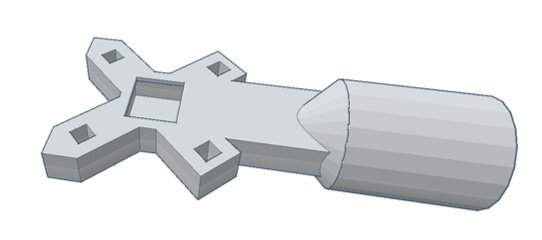}
    \includegraphics[height=2cm]{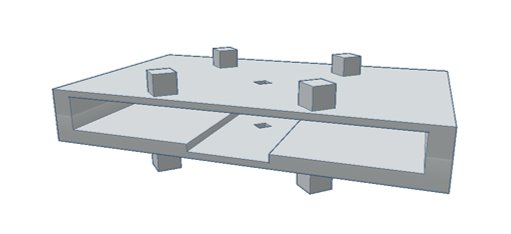}
    \caption{Illustrations of 3D-printed modules for (top) the photomultiplier tubes-waveguide-assembly and (bottom) the Galaxy S6 smartphone. Modules for the S7 and V20 have the same features, but with appropriately altered dimensions.}
    \label{fig:holder}
\end{figure}

To support these detectors and to aid in their alignment, 3D-printed modules were designed for both the phones and the PMT assemblies, illustrated in \Cref{fig:holder}. Square pegs and holes anchor these modules in a way that minimizes the separation between phones and scintillators (\SI{2.4}{mm} of plastic),  maximizing the solid angle sampled.

\subsection{Scintillator tags}

The PMT output is directed to a custom analog front-end connected to an Arduino micro-controller, which functions as a remotely adjustable discriminator. Its two primary roles, to set the three PMT thresholds and to register and record tags, are both driven over a serial connection by a Raspberry Pi computer. Tags registered by the Arduino are saved and converted to Raspberry Pi wall clock time for comparison with the phone data. 

Thresholds for each scintillator assembly are chosen to maximize muon detection efficiency, measured with cosmic muons tagged by other scintillators placed immediately above and below. This measurement quantified the scintillator efficiency only for muons passing through both the top and bottom surfaces of the middle scintillator; the omnidirectional efficiency, which includes muons entering or exiting through the sides, is treated below. For all three scintillators, the efficiency remained close to 100\% well above the chosen thresholds.

\section{CMOS Calibration}

Three phone models were used, which contain the two most common brands of commercial CMOS sensors: Exmor (Sony) and ISOCELL (Samsung).  The phones include two ISOCELL-based Galaxy S6s,  two Exmor-based Galaxy S7s, and two LG V20s, one with each variety of sensor. To record particle hits on the CMOS and perform calibration studies, a dedicated streamlined version of the \textsc{Crayfis} app~\cite{bib:2.65} was developed. This app streams 10-bit RAW-format camera buffers with low frame rates, 2 frames per second (FPS) on the S6s and V20s, and 7 FPS on the S7s, to avoid overburdening the CPU. Individual pixels with digital values above a threshold are considered CMOS {\it hits}; the value of this threshold is varied later to explore the dependence of the efficiency. All hits are saved above the minimum threshold each device is capable of maintaining without overheating.

\begin{figure}
    \centering
    \includegraphics[height=4cm]{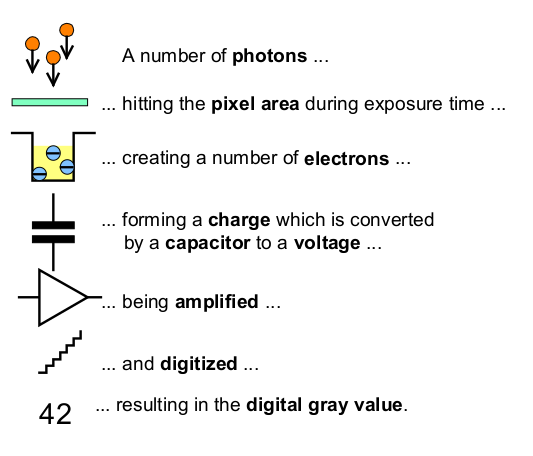}
    \includegraphics[height=4cm]{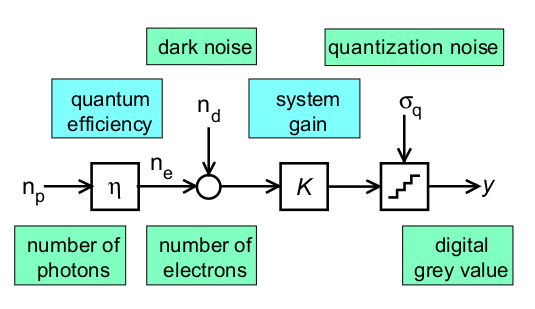}
    \caption{A schematic of the process of photon capture, electron release and conversion to digital pixel values,  from Ref. \cite{bib:6.3}.}
    \label{fig:digi}
\end{figure}

The digital pixel values reported by the smartphone depend on arbitrary and phone-dependent gains. These include spatially-dependent lens-shading gains, which compensate for the decreased intensity of light near the edges of the sensor, further from the aperture; see \Cref{fig:digi}.  To measure efficiencies as a function of the number of photo-electrons collected, this transformation must be deduced and inverted. In addition, the CMOS sensors require masking of hot pixels and synchronization of their clocks with the scintillator readout clock.

\subsection{Determining system gain}

  The per-pixel system gain is not generally accessible from the software.  A technique pioneered in Ref.~\cite{belllabs,10.1117/12.175165} is instead  applied, which relies on Poisson statistics to deduce the system gain from a comparison of pixel means and variances under controlled conditions. 

For a given pixel under fixed exposure conditions, the mean digital pixel value $\mu$ is:
\begin{equation}\label{eq:adc_response}
\mu=K\,(\lambda_e+\lambda_d)
\end{equation}
where $\lambda_e$ and $\lambda_d$ are the mean number of photo-electrons and dark current electrons per exposure, respectively, and $K$ is the unknown system gain. The variance is expressed as:
\begin{equation}
\sigma^2=\sigma_q^2+K^2\,(\sigma_e^2+\sigma_d^2)
\end{equation}
where $\sigma_q^2$, $\sigma_e^2$, and $\sigma_d^2$ are the 
variances of digitization and readout noise, photo-electrons, and dark current electrons, respectively.  Assuming Poisson statistics, this is simply: 
\begin{equation}
\sigma^2=\sigma_q^2+K^2\,(\lambda_e+\lambda_d)
\end{equation}
and inserting $\mu$ from \eqref{eq:adc_response} yields a linear relationship between the mean $\mu$ and variance $\sigma^2$:
\begin{equation}
\sigma^2=\sigma_q^2+K \, \mu \,.
\end{equation}

Both the mean and variance of the digitized pixel values are measured directly, allowing the overall gain $K$ to be deduced by linear regression.  Due to the applied lens-shading gains, $K$ is not constant across the sensor, but varies, typically with a radial symmetry. This dependence is parametrized as:
\begin{equation}\label{eq:Kgain}
K(r)=K_0\,\lambda(r) 
\end{equation}
where 
\begin{equation}
\min_{r} \lambda(r) = 1 \,.
\end{equation}

Controlled light exposures of the phones were performed by securing their rear cameras several inches away from a vertically-oriented white sheet of paper. The lab was illuminated by constant artificial lighting (slightly biased towards longer visible wavelengths) and kept isolated from any additional light sources, natural or artificial. To modulate the exposure, varying thicknesses of white and colored paper were taped over the lenses, constructed in such a way that the means of nearby red, green, and blue channels were comparable. To measure $K(r)$, one phone of each model was exposed to several intensities of light while the exposure time per frame was held constant. Sample means and variances were calculated at each light intensity from runs of 5000 frames each. Linear regressions on each individual pixel were performed, using sample means sufficiently far from both the black level and saturation. A linear fit describes the data well, with $R^2 > 0.985$ for over 99\% of pixels. 

\begin{figure}
    \centering
    \includegraphics[width=\linewidth]{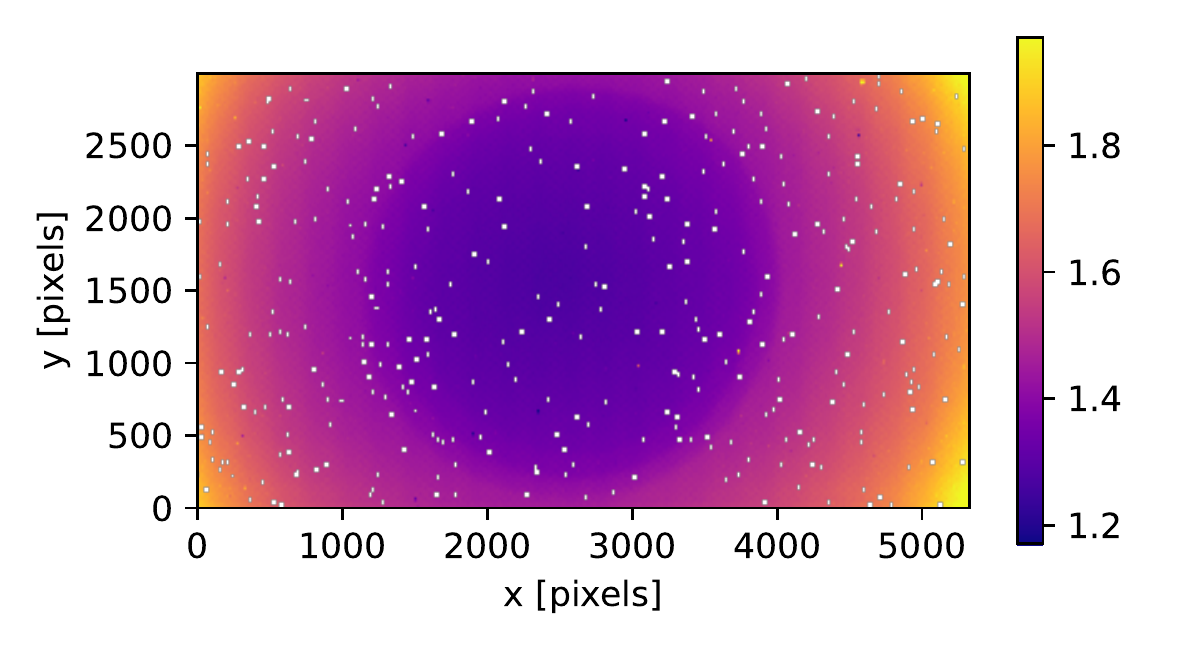}
    \caption{ The spatial dependence of the gain factor $K(x,y)$ over a CMOS sensor, showing a clear lens-shading effect with a radial symmetry.}
    \label{fig:gain}
\end{figure}

The spatial variation of the gain shows a clear radial symmetry for the S6 and S7 devices; see \Cref{fig:gain}. The radial dependence is isolated using 4×4 blocks of downsampled pixels, and parameterized using a piece-wise linear fit, which compares well to the simple corrections needed for a lens-less aperture. The overall constant $K_0$ can then be extracted from pixel counts which have been scaled by $1/\lambda(r)$. Lens-shading gains can be disabled by the user on the V20 devices, such that $\lambda_\mathrm{V20}(r)=1$.

\begin{figure}
    \centering
    \includegraphics[width=0.8\linewidth]{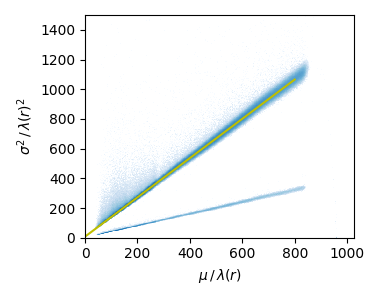}
    \includegraphics[width=0.8\linewidth]{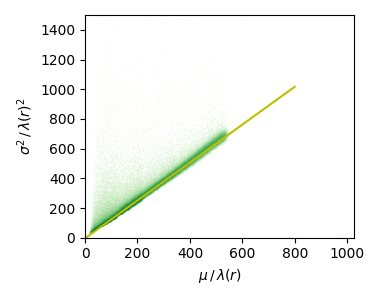}
    \caption{ Pixel sample mean ($\mu$) versus variance ($\sigma^2$), after correcting for lens-shading factor $\lambda(r)$; see text for details. Top shows data from the ISOCELL-sensor LG V20, bottom for the Galaxy S6, colored according to the color channel. A secondary line composed of autofocus pixels is visible for the LG V20. The overall gain factor $K_0$ is the measured slope of a fit (solid line) to the mean-variance relationship.}
    \label{fig:photon_tranfer_linearized}
\end{figure}

A linearized photon transfer curve is constructed by dividing the pixel means by $\lambda(r)$ and the variances by $\lambda(r)^2$.  \Cref{fig:photon_tranfer_linearized} shows these photon transfer curves for two superimposed runs each of (a) a Galaxy S6 and (b) a LG V20, taken with the same procedure as the previous runs. Secondary low-variance curves can be seen in several plots, and are composed of special-purpose autofocus pixels; as these pixels are partially masked, their reported value is instead derived from an average of their neighbors, thereby decreasing their variances. Linear fits were found for each channel in the  color filter array with a Hough transformation; under this approach, outliers such as autofocus pixels and pixels with statistics skewed by saturation were ignored. The gain in electrons per adjusted count can then be read as the slope. For the V20 and S7, the per-channel gains were indistinguishable, while a roughly 5\% discrepancy was found in the S6 color-channel gains.
 
\subsection{Hot-pixel masking}

Some fraction of CMOS pixels are grossly defective and attain values near saturation on nearly every frame, while others have small defects that increase their typical response. Masking these pixels allows for a lower threshold on the number of photo-electrons collected (yielding a higher efficiency) while maintaining the same pass rate.  

\begin{figure}
    \centering
    \includegraphics[width=0.9\linewidth]{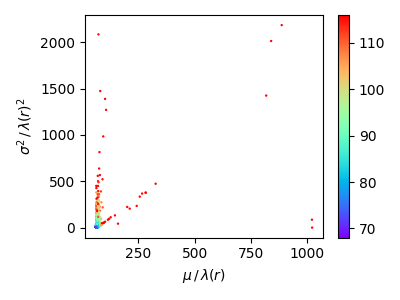}
    \includegraphics[width=0.9\linewidth]{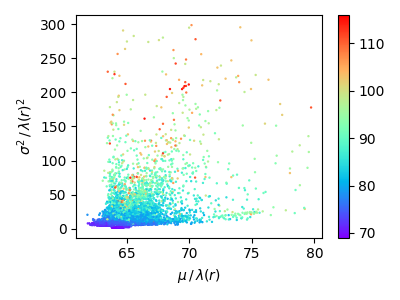}
    \caption{Mean and variance of pixel values, after adjusting for lens-shading gains as described in text, in long dark runs for a random subset of pixels. Before hot-cell masking (top), many outliers are obvious. When removed, (bottom), the distributions are much more compact. The color axis indicates the second highest value recorded by each pixel in 20,000 frames. }
    \label{fig:hot_pixelstats}
\end{figure}

\begin{figure}
    \centering
    \includegraphics[width=0.9\linewidth]{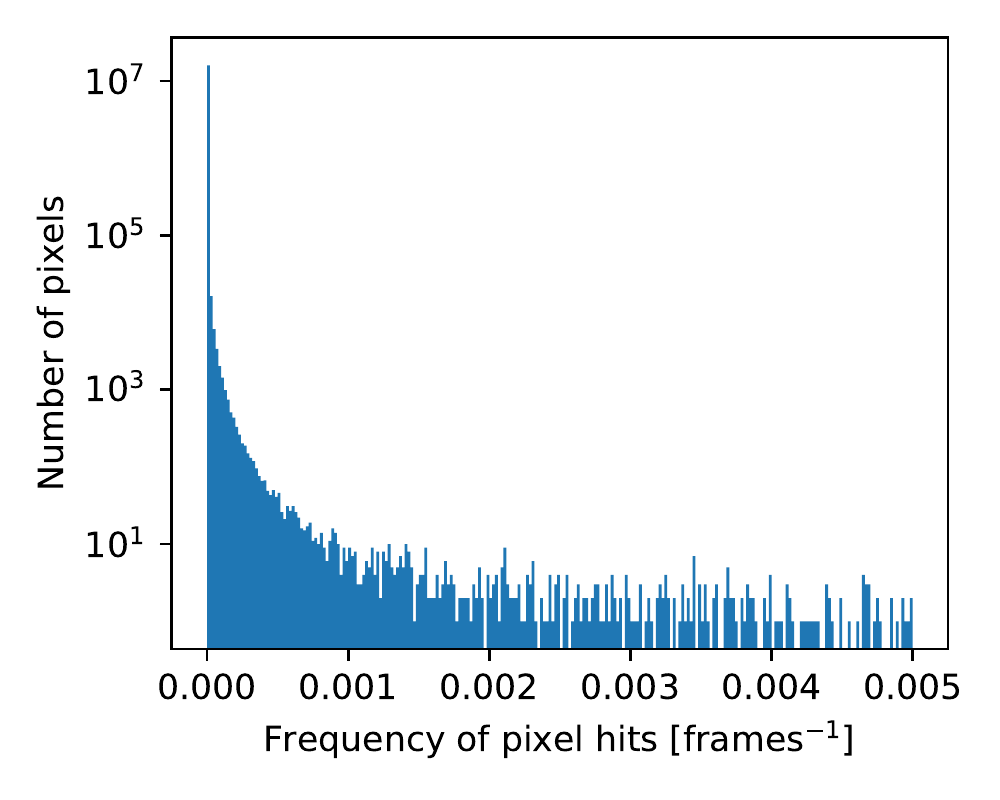}
    \caption{ Hot pixels are identified by examining the frequency at which pixels are above the minimum hit threshold in a long dark run. After removal of extreme outliers as described in \Cref{fig:hot_pixelstats}, the pixel hit rates show a large quiet fraction with close to zero, and a long tail of hot pixels.}
    \label{fig:hot_cosmics}
\end{figure}

First, extreme outliers in mean, variance, and second-highest value are masked; see \Cref{fig:hot_pixelstats}.  Additional hot pixels are identified by examining the rate at which pixels surpass the digital threshold in long, dark exposures; see \Cref{fig:hot_cosmics}. Hot pixels are defined as those which are above the minimum hit threshold most frequently, with a threshold that varies by device to balance reducing background and maintaining muon efficiency.
 
\subsection{Clock Synchronization}

A comparison of scintillator tags and CMOS hits requires synchronization of the clocks used to record them. Though CMOS frame times are provided with nanosecond-level precision, the precisely measured relative frame intervals are not absolutely calibrated to wall clock time, and cannot determine hit time within the frame duration.

\begin{figure}
    \centering
    \includegraphics[width=0.9\linewidth]{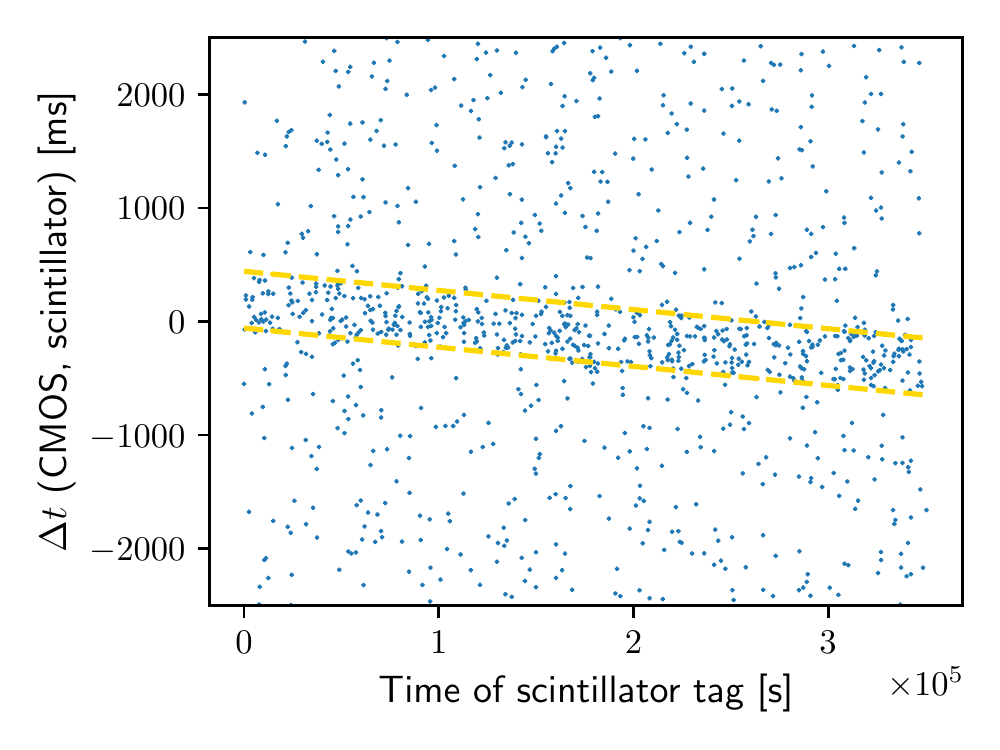}
    \caption{ Time difference between scintillator tags and CMOS hits versus scintillator tag time shows a clear correlation as well as a time-dependent clock discrepancy. Dotted lines show a band fitted via a Hough transform. }
    \label{fig:hodo_drift}
\end{figure}

\begin{figure}
    \centering
    \includegraphics[width=0.95\linewidth]{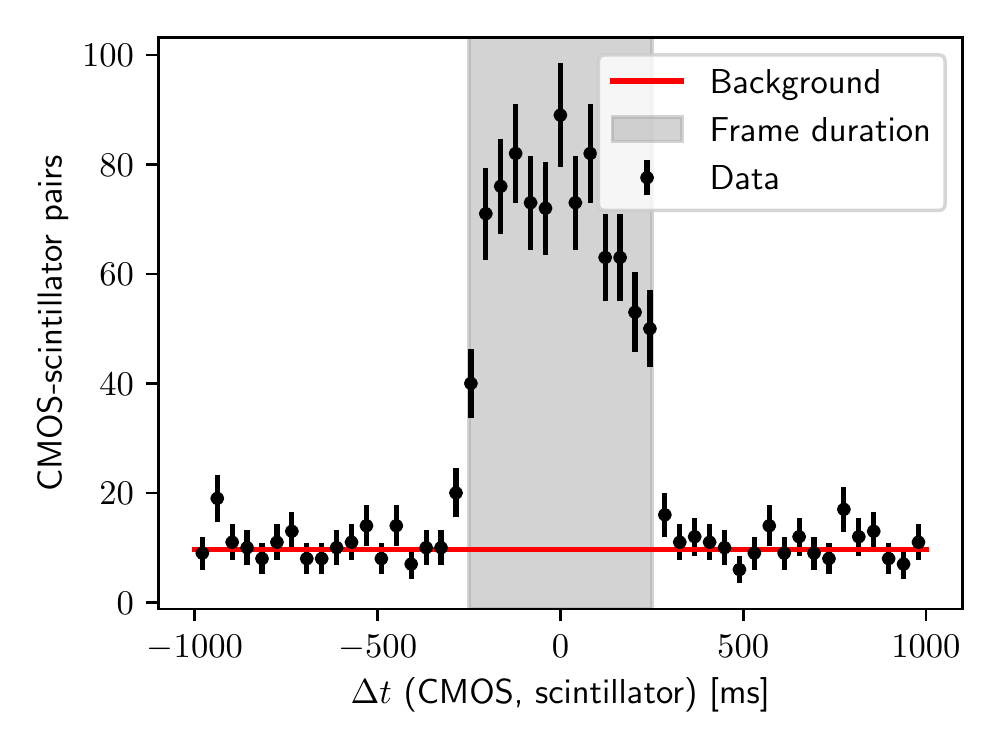}
    \caption{ Difference between scintillator tag and CMOS hit times (black points) shows a band around zero of clear excess over random coincidences (solid red line).}
    \label{fig:hodo_dt_hist}
\end{figure}

To determine the needed correction, the time differences between all possible pairs of scintillator tags and CMOS hits are examined as a function of the scintillator clock time. In \Cref{fig:hodo_drift}, a band of CMOS-scintillator coincidences is clearly visible, with a slope on the order of \SI{1}{\micro s/s}, in agreement with Ref. \cite{drift}. After correction for the slope, the time difference shows a clear excess band near zero; see \Cref{fig:hodo_dt_hist}. The band's finite width reflects the timing resolution of the experiment, which is determined almost entirely by the CMOS frame duration.  

\section{Statistical Analysis}

Measurement of the CMOS efficiency requires accounting for all possible sources of hits, including cosmic muons, secondaries, background radiation, and electronic noise.

Scintillator tags are assumed to be exclusively caused by cosmic muons. The rate of scintillator tags due to electronic and combinatorial noise was measured in a 6-hour run with the scintillators apart and set to the the lowest viable thresholds. Only a single tag was registered among the three possible scintillator pairs, suggesting that tags from noise are negligible. 

Under this assumption, the CMOS muon efficiency $\epsilon$ for a given hit threshold is defined as
\begin{equation}
\epsilon=P(H\:|\:T \cap C)
\end{equation}

\noindent where $H$ is the condition that the CMOS pixel records a hit due to an incident particle,  $C$ is the condition that a particle is incident on the CMOS sensor, and $T$ is the condition that the incident particle is a muon identified by a scintillator tag.

As $H \subset C$, this becomes:
\begin{equation}\label{eq:hodo_eff_frac}
\epsilon = \frac{P(H \: | \: T)}{\alpha}
\end{equation}
where $\alpha = P(C \: | \: T)$ is the geometric acceptance of the CMOS sensor for muons with scintillator tags. The acceptance is estimated through a Monte Carlo simulation of muon trajectories through the experiment. The numerator $P(H \: | \: T)$ is determined entirely by the observed scintillator-CMOS coincidence rates.  

\subsection{Counting coincidences}

To extract $P(H \: | \: T)=\epsilon\,\alpha$ from muons with scintillator tags requires accounting for the combinatorial background that contributes to CMOS hits. 

Each scintillator tag is associated with a set of CMOS frames within corrected relative time $\tau/2$, where $\tau$ is fixed at 120\% of the frame duration; this set is referred to as a {\it block}. To avoid the complication of multiple coincident scintillator tags, only blocks in which each frame is associated to a single tag are kept. The probability that a block with $n$ frames contains at least one CMOS frame with a hit is given by:
\begin{equation}\label{eq:block_prob}
p_n=1 - (1-\epsilon \alpha)(1-\eta)^n
\end{equation}
where $\eta$ is the probability that a frame contains a hit due to something other than a tagged muon, such as sensor noise, ambient radiation, or muons outside the detector's acceptance.
If $B_n$ is the total number of blocks with $n$ frames, then the number of those blocks containing at least one CMOS hit ($B_{H,n}$) follows the binomial distribution:
\begin{equation}\label{eq:binom}
B_{H,n} \sim \mathrm{Binom}(B_n, p_n)
\end{equation}
allowing $p_n$ to be estimated as:
\begin{equation}\label{eq:binom_pn}
\begin{aligned}
\widehat{p}_\mathrm{n} &= \frac{B_{H,n}}{B_{n}} \,, \\
\widehat{\sigma}_{p_n}^2 &= \frac{\widehat{p}_n(1-\widehat{p}_n)}{B_n} \,.
\end{aligned}
\end{equation}
The number of untagged frames, $U$, and the subset with a CMOS hit, $U_H$, are used to measure $\eta$, \eqref{eq:binom_pn}:
\begin{equation}
\begin{aligned}
\widehat{\eta} &=\frac{U_H}{U} \,, \\
\widehat{\sigma}_{\eta}^2 &=\frac{\widehat{\eta}(1-\widehat{\eta})}{U} \,.
\end{aligned}
\end{equation}

An estimate for $\epsilon \alpha$ can then be calculated as a weighted average over the unique values of $n$:
\begin{equation}
\widehat{\epsilon \alpha}=1- \left(\frac{1}{\sum_n B_n}\right) \cdot \sum_n \frac{B_n(1-\widehat{p}_n)}{(1-\widehat{\eta})^n} 
\end{equation}
\begin{equation}
\label{eq:var_ea}
\begin{split}
\widehat{\sigma}_{\epsilon \alpha}^2 =\left(\frac{1}{\sum_n B_n}\right)^2 \cdot \left[\sum_n \frac{\widehat{p}_n}{B_n(1-\widehat{p}_n)}\left(\frac{B_n(1-\widehat{p}_n)}{(1-\widehat{\eta})^n}\right)^2 \right. \\
\left. +\frac{\widehat{\eta}}{U(1-\widehat{\eta})}\left(\sum_n\frac{n B_n(1-\widehat{p}_n)}{(1-\widehat{\eta})^n}\right)^2\right] \,.
\end{split}
\end{equation}

\noindent Expressions were validated using toy Monte Carlo  with a variety of hit rates, noise levels and coincidence window sizes.

\subsection{Geometrical acceptance}

\begin{figure}
    \centering
    \includegraphics[width=0.9\linewidth]{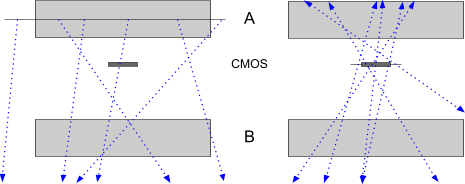}
    \caption{Illustrations of the toy Monte Carlo used to measure the geometric acceptance, in which simulated muons are generated from (left) scintillator A, and (right) the CMOS.}
    \label{fig:diagram_mc}
\end{figure}

The geometrical acceptance $\alpha$ of the CMOS sensor is calculated using Monte Carlo simulations of the experimental configuration. Simulated muons are generated uniformly on planes bisecting the scintillators and CMOS 
with a directional intensity:
\begin{equation}
\label{eq:zenith}
I(\theta)=I_0\cos^n\theta
\end{equation} 
where we estimate $n=2.0\pm0.1$ based on Figure 3.60 of Ref. \cite{grieder}.
Ray-tracing is used to compute the relevant acceptances.
Through Bayes' theorem, the value of $\alpha = P(\mathrm{CMOS} \: | \: A \cap B)$ is obtained from the probabilities ${P(A \cap B \: | \: \mathrm{CMOS})}$ and ${P(B \: | \: A)}$, requiring the left and right geometries in \Cref{fig:diagram_mc}, respectively.

To set the CMOS position, a representative device of each model was disassembled to find the location of the CMOS chip within the phone. By comparing the dimensions of the individual modules to those of the total assembly, the average gap size between the modules was found to be \SI{0.4}{mm}. These gaps were included in the geometry and a \SI{0.4}{mm} uncertainty was  assigned to all distances within the simulation. 

Another source of uncertainty in the geometric acceptance is the rate at which muons only partially traversing the scintillator deposit enough energy to pass the PMT threshold. This rate is estimated by varying the distance between the physical scintillators and the PMT thresholds, which both modify the contribution from partially-traversing muons. This is parametrized in the Monte Carlo by a threshold $L$ on the muon path length through the scintillator volumes, with a best-fit value of $3.7 \pm 0.3\,\si{mm}$ from the observed coincidence rates assuming $n=2$ in \eqref{eq:zenith}.

To propagate these uncertainties to the computed geometric acceptance, each simulation was repeated 100 times with independent Gaussian displacements in the measured quantities: the phone and scintillator positions and the path length threshold $L$. The value and uncertainty of the acceptance are estimated from the mean and standard deviation of these trials. To estimate the additional uncertainty from the zenith angle distribution, this analysis was repeated for $n=1.9$ and $n=2.1$, with separate fits for $L$. 
A summary of the uncertainties are shown in \Cref{tab:uncertainty}.

As a final cross-check of the geometric acceptance, our measurement of the observed cosmic muon flux is compared to  well-established values. 
A lead brick measuring $8"\times 4"\times 2"$ was suspended several millimeters above the  scintillators, with the 4" side extending vertically to shield the experiment from cosmic ray photons, electrons, and positrons. Our measured cosmic muon flux is $J_1=141\pm 5\,\si{Hz/m^2}$, which agrees well with Ref. \cite{grieder}.

\begin{table}[]
    \centering
        \caption{Estimates of the uncertainty in muon efficiency for an S7 smartphone for various thresholds on $N_e$, the number of photo-electrons collected.  Statistical uncertainty
        (Eq. \eqref{eq:var_ea}) and systematic uncertainty due to experimental alignment, partially-traversing muons, and the zenith-angle distribution (Eq. \eqref{eq:zenith}) are provided.}
    \begin{tabular}{c|r|r|r}
            \hline
            \hline
            \multicolumn{4}{c}{Uncertainty in efficiency}\\
            \hline
            Contribution & $N_e=20$ & $N_e=50$ & $N_e=100$ \\
        \hline
        Statistical         & $.039$ & $.023$ & $.019$ \\
        Exper. alignment    & $.041$ & $.034$ & $.020$ \\
        Partial trav. muons & $.022$ & $.018$ & $.011$ \\
        Zenith angle distr. & $.025$ & $.021$ & $.012$ \\
        \hline
        Total & $.067$ & $.051$ & $.033$ \\
                \hline
                                \hline
    \end{tabular}
    \label{tab:uncertainty}
\end{table}

\section{Results and discussion}

\begin{figure}
    \centering
    \includegraphics[width=0.9\linewidth]{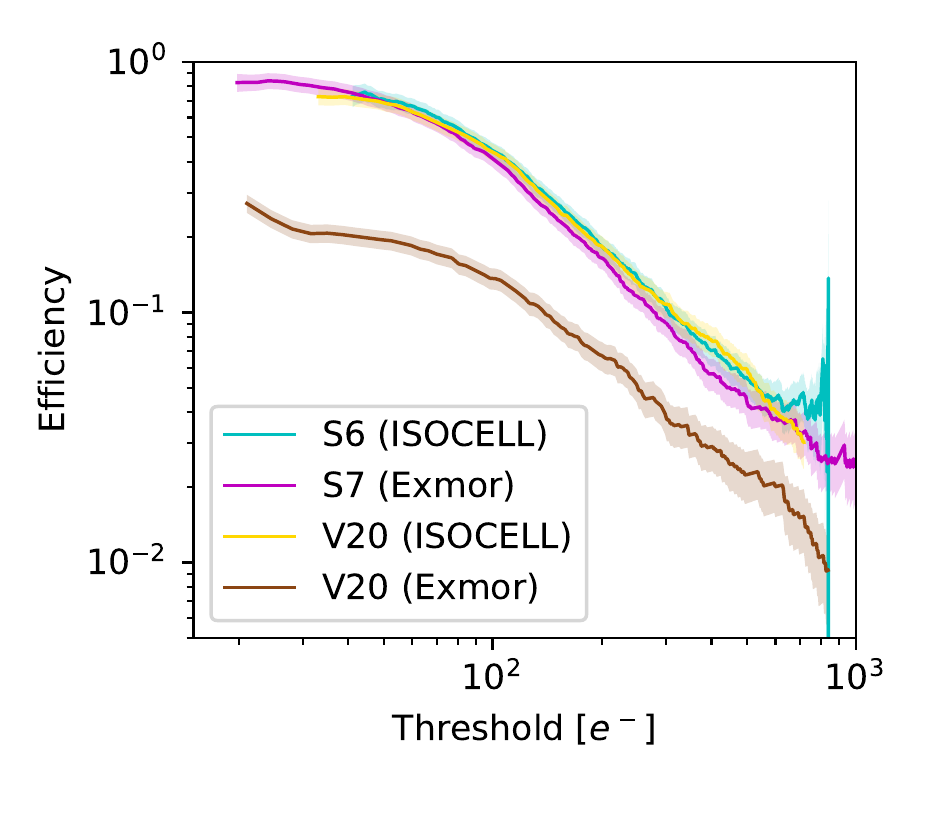}
    \caption{CMOS hit efficiencies for cosmic ray muons  versus threshold on collected photo-electrons, with uncertainties shown as shaded bands, for four different phone models. The top S6 and bottom S7 are plotted with both LG V20s.}
    \label{fig:hodo_efficiency_all}
\end{figure}

Efficiency measurements were made from two-week lead-shielded runs of each phone.  The measured efficiencies as a function of a threshold on the number of collected photo-electrons are shown in \Cref{fig:hodo_efficiency_all} and given in Table~\ref{tab:res}. Note that the errors across each curve are highly correlated: as the threshold increases, smaller subsets of the same dataset are used. Systematic errors in $\alpha$ scale each curve as a whole. 

Nearly identical efficiencies were found in the pairs of S6s and S7s; hence only one of each is shown in \Cref{fig:hodo_efficiency_all}. By contrast, the two sensors in the V20s, the IMX298 (Exmor) and S5K2P7 (ISOCELL), exhibited strikingly different performances. The anomalous behavior of the IMX298 is almost certainly caused by its dynamic hot pixel correction~\cite{bib:imx298}; though the details of the algorithm are proprietary, it appears to mask isolated bright pixels, including muon tracks. Accordingly, muons tagged by the outermost scintillators, having smaller zenith angles and consequently exciting fewer pixels, were triggered at roughly half the efficiency of the full muon sample. 




For all but the LG V20 with the Exmor sensor, the values of the effective area at the lowest feasible thresholds are within the range considered by Ref. \cite{bib:2.65}, falling between $(1.4$ -- $2.2)\times 10^{-5}$\si{m^{-2}}, when the minimal acceptance $A=A_\mathrm{CMOS}$ is used. 

\begin{table}[h!]
    \centering
        \caption{Measured efficiencies for several phone models, as a function of a threshold on the number of photo-electrons. The first column is the minimum viable threshold, also given, below which the sensor noise and corresponding uncertainties rapidly increase. Uncertainties include all statistical and systematic uncertainties, dominated by the uncertainty on the geometric acceptance. For the two S6s and S7s, the average values of the (nearly identical) means and uncertainties are provided. }
    \begin{tabular}{c|r|r|r}
            \hline
            \hline
            \multicolumn{4}{c}{Selected efficiencies}\\
            \hline
        Phone       & Min threshold & $N_e=50$ & $N_e=100$ \\
        \hline
        S6          &  $.74\pm.07$ @ 43$e$ &    $.72\pm.06$ & $.47\pm.04$ \\
        S7          &  $.87\pm.07$ @ 20$e$ & $.70\pm.05$ & $.43\pm.03$ \\
        V20 ISOCELL & $.73\pm.06$ @ 33$e$  & $.69\pm.05$ & $.44\pm.03$ \\
        V20 Exmor &  $.27\pm.02$ @ 21$e$ & $.19\pm.02$ & $.14\pm.01$ \\
                \hline
                                \hline
    \end{tabular}
    \label{tab:res}
\end{table}

Our results directly demonstrate that the built-in capabilities of readily-available smartphones are sufficient for the detection of cosmic-ray muons with high efficiency, and are, to our knowledge, the first to quantify the CMOS sensor efficiency which can be achieved.

\section{Acknowledgements}

JS was supported by a generous grant from the Jenkins Family Foundation.  The authors are grateful to Mauricio and John Paul Vela for their contributions to building the experiment, and to Bob Hirosky and Eric Albin for helpful comments on earlier drafts.

\bibliography{sensor}

\end{document}